# Superconductivity at 52 K in iron-based F-doped layered quaternary compound Pr[O$_{1-x}$F$_x$]FeAs

Zhi-An Ren*, Jie Yang, Wei Lu, Wei Yi, Guang-Can Che, Xiao-Li Dong, Li-Ling Sun, Zhong-Xian Zhao*

National Laboratory for Superconductivity, Institute of Physics and Beijing National Laboratory for Condensed Matter Physics, Chinese Academy of Sciences, P. O. Box 603, Beijing 100190, P. R. China

(Dated: March 28, 2008)

Since the discovery of copper oxide superconductor in 1986 [1], extensive efforts have been devoted to the search of new high-$T_c$ superconducting materials, especially high-$T_c$ systems other than cuprates. The recently discovered quaternary superconductor La[O$_{1-x}$F$_x$]FeAs with the superconducting critical transition $T_c$ of 26 K [2], which has a much simple layered structure compared with cuprates, has attracted quick enthusiasm and is going to become a new high-$T_c$ system [3-6]. Here we report the discovery of bulk superconductivity in the praseodymium-arsenide oxides Pr[O$_{1-x}$F$_x$]FeAs with an onset drop of resistivity as high as 52 K, and the unambiguous zero-resistivity and Meissner transition at low temperature, which will place these quaternary compounds to another high-$T_c$ superconducting system explicitly.

The superconducting Pr[O$_{1-x}$F$_x$]FeAs samples were prepared by a high pressure synthesis method. Fine powders of PrAs (pre-synthesized by Pr pieces and As powder), Fe, Fe$_2$O$_3$, FeF$_3$ (the purities of all starting chemicals are better than 99.99%) were mixed together according to the stoichiometric ratio of Pr[O$_{1-x}$F$_x$]FeAs, then ground thoroughly and pressed into small pellets. The pellets with the nominal composition of Pr[O$_{0.89}$F$_{0.11}$]FeAs were sintered in a high pressure synthesis apparatus under a pressure of 6 GPa and temperature of 1250$^{\circ}$C for 2 hours. Here we note that the samples usually cannot fully react during the short high-pressure synthesis time, nevertheless this synthesis method has a special merit that can reduce the fluorine loss greatly by press-seal compared with the vacuum-quartz tube seal method that commonly adopted [2]. The structure of the samples was characterized by powder X-ray diffraction (XRD) on an MXP18A-HF-type diffractometer with Cu-K$_\alpha$ radiation from 20$^{\circ}$ to 80$^{\circ}$ with a step of 0.01$^{\circ}$.

The XRD patterns of all samples indicate the main phase of the PrOFeAs structure (PDF card, no. 52-0412, S.G. = P4/nmm, a = 3.9853 Å, c= 8.595 Å) with shrunk lattice. One of them that studied in this paper (with the highest $T_c$) was plotted in Fig. 1, with the lattice parameters a = 3.967(1) Å and c = 8.561(3) Å calculated by the least-square fit method, which shows a shrinkage of 0.45% for a-axis and 0.40% for c-axis that caused by the substitution of $O^{2-}$ by $F^-$. The calculated XRD peaks were displayed as vertical bars for comparison. Slight impurity peaks were marked with '*', which are mainly from un-reacted residuals.

The resistivity was measured by the standard four-probe method in a physical property measurement system (Quantum Design, PPMS) under magnetic fields of 0 and 9 T. The results are shown in Fig. 2. The zero-field resistivity shows a clear drop as the temperature down to 52 K, and becomes unmeasurable at 44 K. The middle of the superconducting transition is 47 K. Upon the magnetic field of 9 T, the zero transition is shifted to 35 K, while the onset transition has no obvious change, which may indicate an extra large upper critical field of this new superconductor.

The magnetization measurements were performed on a Quantum Design MPMS XL-1 system during warming cycle under fixed magnetic field after zero field cooling (ZFC) or field cooling (FC) process. The AC-susceptibility data (with the measuring frequency of 997.3 Hz and the amplitude of 0.5 Oe) and DC-susceptibility data (measured under a magnetic field of 1 Oe) are shown in Fig. 3 (we note that the negative background on DC curves is caused by the magnetic residuals, which can be eliminated by raising synthesis temperature and time, but this will decrease $T_c$ due to the loss of fluorine; the details will be reported later). The 10% and 90% transition on the ZFC curve is 43.5K and 38K respectively. The sharp magnetic transition indicates the good quality of this superconductor. The onset diamagnetic transition determined from the differential ZFC curve is near 50 K. The diamagnetic shielding signal at 5 K is about 90%, which indicate the bulk nature of this new $Pr[O_{0.89}F_{0.11}]FeAs$ superconductor.

Acknowledgements:

We thank Prof. Hai-Hu Wen and Mr. Gang Mu for their kind helps in resistivity measurements. This work is supported by Natural Science Foundation of China (NSFC) and 973 program of China (No. 2007CB925002). We also acknowledge the support from EC under the project COMEPHS TTC.

Corresponding Authors:

Zhi-An Ren: renzhian@aphy.iphy.ac.cn

Zhong-Xian Zhao: zhxzhao@aphy.iphy.ac.cn

References:

[1]. Bednorz, J. G. and Muller, K. A. Possible high $T_c$ superconductivity in the barium-lanthanum-copper-oxygen system. *Z. Phys. B.* **64**, 189 (1986).

[2]. Kamihara Y., Watanabe T., Hirano M. and Hosono H. Iron-Based Layered Superconductor La[$O_{1-x}F_x$]FeAs$_x$ (x = 0.05-0.12) with $T_c$ = 26 K. *J. Am. Chem. Soc.* **130,** 3296 (2008).

[3]. Wen, H. H., Mu, G., Fang, L., Yang, H., Zhu, X. Y. Superconductivity at 25 K in hole doped $La_{1-x}Sr_x$OFeAs. *Condmat:arXiv*, 0803-3021 (2008).

[4]. Chen, X. H., Wu, T., Wu, G., Liu, R. H., Chen, H., Fang, D. F. Superconductivity at 43 K in Samarium-arsenide Oxides SmFeAsO$_{1-x}$F$_x$. *Condmat:arXiv*, 0803-3603 (2008).

[5]. Chen, G. F., Li, Z., Wu, D., Li, G., Hu, W. Z., Dong, J., Zheng, P., Luo, J. L., Wang, N. L. Superconductivity at 41 K and its competition with spin-density-wave instability in layered CeO$_{1-x}$F$_x$FeAs. *Condmat:arXiv*, 0803- 3790 (2008).

[6]. Lu, W., Yang, J., Dong, X. L., Ren, Z. A., Che, G. C., and Zhao, Z. X. Pressure Effect on the superconducting properties of LaO$_{1-x}$F$_x$FeAs (x = 0.11) superconductor. *Condmat:arXiv*, 0803-4266 (2008).

Figure captions:

Figure 1: X-ray powder diffraction pattern of the Pr[$O_{0.89}F_{0.11}$]FeAs superconductor, '*' indicates impurity phases; the vertical bars correspond to the calculated diffraction peaks based on the experimental lattice parameters.

Figure 2: The temperature dependence of resistivity for the Pr[$O_{0.89}F_{0.11}$]FeAs superconductor.

Figure 3: The temperature dependence of AC-susceptibility, DC-susceptibility, and differential ZFC curve for the Pr[$O_{0.89}F_{0.11}$]FeAs superconductor.

Figure 1:

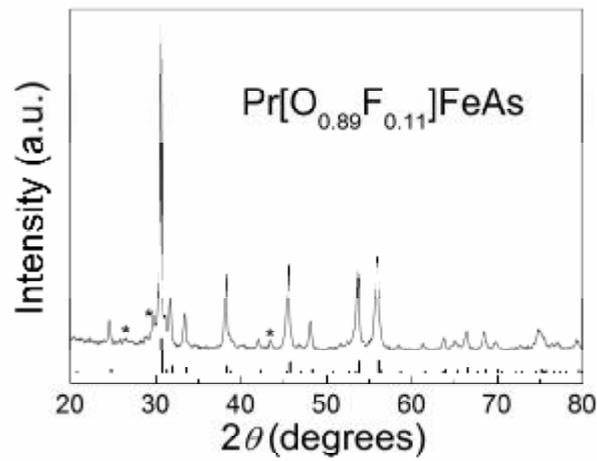

Figure 2:

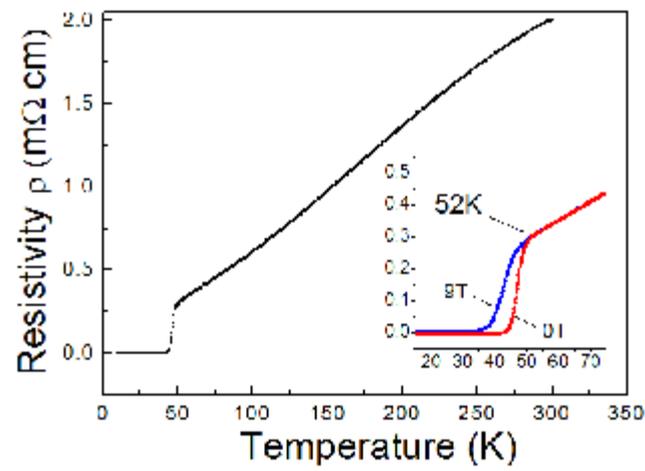

Figure 3:

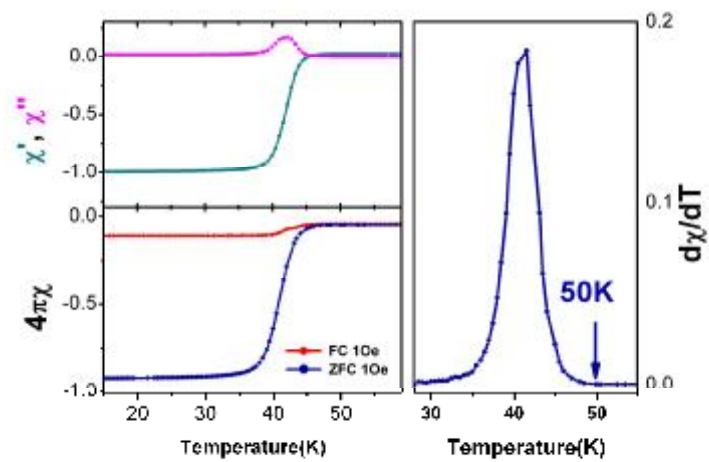